\documentstyle{article}

\begin{document}
\title{Continuity equations for entanglement}
\date{}
\author{Pedro Sancho \\ Centro de L\'aseres Pulsados CLPU \\ Parque Cient\'{\i}fico, 37085 Villamayor, Salamanca, Spain}

\maketitle
\begin{abstract}
We introduce a complex purity density and its associated current for
pure states of continuous variable systems. The scheme is
constructed by analogy with the notions of probability density and
probability current. Taking advantage of the formal continuity
equations obtained this way we can introduce an entanglement
subdynamics. We suggest the use of the dimensionality of this
subdynamics as a potential measure of the complexity of
entanglement. The scheme also provides insights into the relation
between the global and local aspects of quantum correlations.
\end{abstract}

\section{Introduction.}

An elegant and intuitive description of the behavior of
probability in quantum theory is provided by the continuity equation
relating probability density and its associated probability current
\cite{Mes,Gal}. This formulation has been used some times to draw
formal analogies with the theory of fluids.

The reason behind a continuity equation is the conservation of a
physical magnitude, in the previous case the probability. It is
well-known that the entanglement of a system of freely evolving
particles is also conserved. Then it is natural to consider what
type of continuity equation is associated with this conservation
law.

We address here the problem for continuous variable systems. In the
case of pure states, the only one we shall consider in the paper,
the purity is an entanglement measure. It is a measure well-suited for these
systems and we shall concentrate on it. The Schmidt number, the
inverse of the purity, has been frequently used in the literature
\cite{Ebe,Fed}. Concurrence \cite{Ple}, another popular entanglement measure,
is also directly related to purity \cite{pr}. We derive the purity
density, which is complex, showing that we must deal with a pair of
real densities. Similarly, in our scheme we have two currents
instead of one, as it was the case for probability. These densities
and currents are connected by a pair of continuity equations.

The above equations are also interesting when we move to the realm
of interacting particles. In this scenario the interaction potential
connects the behavior of the real and imaginary parts of the
densities and currents. The imaginary part of the density is
contained in the source term of the real continuity equation and
viceversa. We can identify the creation and destruction of
entanglement with these sources terms.

The formalism of continuity equations provides a solid basis to
define the underlying dynamics of entanglement or entanglement
subdynamics. It is based on the temporal evolution of the density
and the current. We can use the dimensionality of this subdynamics to
introduce a measure of the complexity of the entanglement
phenomenon. Moreover, the existence of this subdynamics suggests a
picture where the entanglement phenomenon is continuous (or local)
in the dynamical space, a character that is lost (it becomes global)
when we move to the physical space.

We are only aware of a previous work where the existence of laws of
entanglement conservation has been considered \cite{Sou}. These
authors also use the purity as measure of the entanglement degree
but they consider   discrete variables instead of continuous ones.
Moreover, the two parties in \cite{Sou} are continuously
interacting, explicitly excluding free evolution scenarios. The
analysis in \cite{Sou} focuses on the search of invariant
quantities, which can be obtained by the introduction of a third
party.

The conservation of information in quantum physics has also been
considered in the literature \cite{ron}, raising some questions
about conserved quantities and information flows that in some
aspects resemble the discussion presented here.

\section{Purity}

In this section we briefly review the basic expressions for the
purity as an entanglement measure. The purity, a simple tool to
distinguish between pure states and mixtures, can be also used as an
entanglement measure of pure states (in this paper we shall not
consider mixtures). It is given by
\begin{equation}
\Pi= Tr _{\bf x} (\hat{\rho}_{\bf x}^2)= Tr _{\bf y} (\hat{\rho}_{\bf y}^2)
\end{equation}
where we have used the reduced density matrices
\begin{equation}
\hat{\rho}_{\bf x} =Tr _{\bf y} (|\psi ><\psi |)  \; ; \;  \hat{\rho}_{\bf y} =Tr _{\bf x} (|\psi ><\psi |)
\end{equation}
We only consider the two-particle case, denoting by {\bf x} and {\bf
y} the spatial coordinates of the particles. In the above equations
$Tr _{\mu}$ denotes the trace with respect to the variable $\mu$.
On the other hand, $|\psi >$ is the state of the system. In the position representation it can be expressed as
\begin{equation}
|\psi >= \int d^3{\bf x} \int d^3 {\bf y} \psi ({\bf x},{\bf y}) |{\bf x}>|{\bf y}>
\end{equation}
with $\psi$ the wavefunction of the system. $\psi ({\bf x},{\bf y})$ determines
the amplitude of the mode $|{\bf x}>|{\bf y}>$ in the state
$|\psi>$.

In terms of the wave functions the purity is
\begin{equation}
\Pi=\int d^3{\bf x} \int d^3{\bf y} \int d^3{\bf X} \int d^3{\bf Y} \psi ^* ({\bf x},{\bf y}) \psi ({\bf X},{\bf y}) \psi ^* ({\bf X},{\bf Y}) \psi ({\bf x},{\bf Y})
\label{eq:ss}
\end{equation}
The Schmidt number, its inverse, is frequently used in the
literature. We note that in a similar way we could also use the
concurrence. It is related to the purity by the simple expression
${\cal C} = \sqrt{2(1-\Pi)}$ \cite{pr}. Due to the squared root of
the integral it is a little bit harder to handle than the purity,
but all the conclusions obtained for $\Pi$ can be easily translated
to ${\it C}$.

\section{The purity density}

As it is well-known entanglement is conserved during free evolution.
In our case this can be verified by a simple calculation. On the
other hand, a law of conservation leads to a continuity equation for
the conserved variable. Let us  study this equation. First of all we
must introduce the density of the variable $\Pi$. From Eq.
(\ref{eq:ss}) we can express $\Pi$ as
\begin{equation}
\Pi=\int d^3{\bf x} \int d^3{\bf y} \int d^3{\bf X} \int d^3{\bf Y} \pi ({\bf x},{\bf y},{\bf X},{\bf Y})
\end{equation}
with
\begin{equation}
\pi ({\bf x},{\bf y},{\bf X},{\bf Y}) =\psi ^* ({\bf x},{\bf y}) \psi ({\bf X},{\bf y}) \psi ^* ({\bf X},{\bf Y}) \psi ({\bf x},{\bf Y})
\end{equation}
Clearly $\pi$ plays the role of $\Pi$-density.

One immediate consequence of its definition is that $\pi$ is a
function defined in $R^{12}$. This clearly differs from the
probability density for a two-particle system, which acts in
$R^{6}$. In both cases they are densities not defined in the
physical space $R^3$. This property differs from the definition of
densities in classical systems as fluids, where the space of the
density is the physical space.

Another relevant property of $\pi$ is that it is, in general, a
complex variable. This can be easily seen from its polar
decomposition. For the wave function the decomposition reads $\psi
({\bf x},{\bf y})=R({\bf x},{\bf y}) e^{i\varphi ({\bf x},{\bf
y})}$, and for $\pi$:
\begin{eqnarray}
\pi ({\bf x},{\bf y},{\bf X},{\bf Y}) =R ({\bf x},{\bf y}) R ({\bf X},{\bf y}) R ({\bf X},{\bf Y}) R ({\bf x},{\bf Y}) \times \nonumber \\
\exp (i(-\varphi ({\bf x},{\bf y}) + \varphi ({\bf X},{\bf y}) -\varphi ({\bf X},{\bf Y})+\varphi ({\bf x},{\bf Y})))
\end{eqnarray}
It is clear that, except for $\varphi $ being a constant function or
other special cases, $\pi$ has an imaginary part.

From a physical point of view a complex density does not make sense.
It must be interpreted as a pair of real densities, $\pi _R$ and $\pi _I$,
with $\pi =\pi _R+i\pi _I$. Thus, there is not one but two densities
associated with $\Pi$.

As $\Pi$ is a real number the total contribution of $\pi _I$ must vanish
\begin{equation}
\int d^3{\bf x} \int d^3{\bf y} \int d^3{\bf X} \int d^3{\bf Y} \pi _I({\bf x},{\bf y},{\bf X},{\bf Y})=0
\label{eq:doz}
\end{equation}
However, as we shall see later, this property does not mean that $\pi _I$ is physically irrelevant for the problem.

We can represent $\pi _R$, $\pi _I$ and $\Pi$ as functions acting between the mathematical spaces
\begin{equation}
\pi _R: R^{12} \rightarrow R   \; ; \;  \pi _I: R^{12} \rightarrow R
\end{equation}
and
\begin{equation}
\Pi :C \rightarrow R
\end{equation}
The composite action of $\pi$ and $\Pi$ is
\begin{equation}
\Pi \circ \pi : R^{12} \rightarrow C \rightarrow R
\end{equation}
with the symbol $\circ $ denoting the consecutive application of the
two functions. This chain shows that the dynamics of purity for a
two-particle system takes place in the mathematical space $ R^{12}$
and implies two real densities. The dimensionality of the problem is
much larger than that of the probability dynamics, which for two
particles is only $R^6$.

\section{Purity current and continuity equations}

With the $\Pi$-density we can associate a complex $\Pi $-current obeying
a continuity equation. The procedure to derive the equation is the
usual one. We take the derivative $\partial \pi/\partial t$ which can
be evaluated using the free two-particle Schr\"{o}dinger equation
\begin{equation}
i\hbar \frac{\partial \psi}{\partial t}({\bf x},{\bf y})=-\frac{\hbar ^2}{2m} \Delta _{\bf x} \psi ({\bf x},{\bf y}) -\frac{\hbar ^2}{2M} \Delta _{\bf y} \psi ({\bf x},{\bf y})
\end{equation}
and its complex conjugate. After simple manipulations we obtain
\begin{equation}
\frac{\partial \pi }{\partial t} + \nabla \cdot {\bf J} =0
\label{eq:dsi}
\end{equation}
The $\nabla $-operator and the current ${\bf J}$ are defined in twelve dimensions. The operator can be expressed as
\begin{equation}
\nabla \equiv (\nabla _{\bf x}, \nabla _{\bf y},\nabla _{\bf X}, \nabla _{\bf Y})
\end{equation}
being $\nabla _{\bf x}$ the usual three-dimensional gradient
operator for the variables ${\bf x}$,....

Similarly, the current has four three-dimensional components
\begin{equation}
{\bf J} \equiv ({\bf J}_{\bf x}, {\bf J}_{\bf y},{\bf J}_{\bf X}, {\bf J}_{\bf Y})
\end{equation}
With this notation we have $\nabla \cdot {\bf J} = \nabla _{\bf x}
\cdot {\bf J}_{\bf x} + \nabla _{\bf y} \cdot {\bf J}_{\bf y}
+\nabla _{\bf X} \cdot {\bf J}_{\bf X} +\nabla _{\bf Y} \cdot {\bf
J}_{\bf Y} $. The explicit expressions for the four components of
the current are
\begin{equation}
{\bf J}_{\bf x}=-\frac{i\hbar}{2m} \psi  ({\bf X},{\bf y}) \psi ^* ({\bf X},{\bf Y})(\psi ({\bf x},{\bf Y}) \nabla _{\bf x} \psi ^* ({\bf x},{\bf y}) - \psi ^* ({\bf x},{\bf y}) \nabla _{\bf x} \psi ({\bf x},{\bf Y}))
\end{equation}

\begin{equation}
{\bf J}_{\bf y}=-\frac{i\hbar}{2M} \psi  ({\bf x},{\bf Y}) \psi ^* ({\bf X},{\bf Y})(\psi ({\bf X},{\bf y}) \nabla _{\bf y} \psi ^* ({\bf x},{\bf y}) - \psi ^* ({\bf x},{\bf y}) \nabla _{\bf y} \psi ({\bf X},{\bf y}))
\end{equation}
\begin{equation}
{\bf J}_{\bf X}=-\frac{i\hbar}{2m} \psi ^*  ({\bf x},{\bf y}) \psi ({\bf x},{\bf Y})(\psi ({\bf X},{\bf y}) \nabla _{\bf X} \psi ^* ({\bf X},{\bf Y}) - \psi ^* ({\bf X},{\bf Y}) \nabla _{\bf X} \psi ({\bf X},{\bf y}))
\end{equation}
and
\begin{equation}
{\bf J}_{\bf Y}=-\frac{i\hbar}{2M} \psi ^* ({\bf x},{\bf y}) \psi ({\bf X},{\bf y})(\psi ({\bf x},{\bf Y}) \nabla _{\bf Y} \psi ^* ({\bf X},{\bf Y}) - \psi ^* ({\bf X},{\bf Y}) \nabla _{\bf Y} \psi ({\bf x},{\bf Y}))
\end{equation}
Clearly ${\bf J}$, as $\pi $, is a complex variable. We can introduce
its decomposition in real and imaginary parts, ${\bf J}={\bf J}_R +
i {\bf J}_I$.

The continuity-like Eq. (\ref{eq:dsi}) is only formal. A complex
density and a complex current do not make any physical sense. They
must be understood as two densities and two currents connected by
two different continuity equations:
\begin{equation}
\frac{\partial \pi _{\xi}}{\partial t} + \nabla \cdot {\bf J}_{\xi} =0
\end{equation}
with $\xi =R,I$.

\section{Interacting particles}

After deriving the continuity equations for free evolving particles
we must analyze how they are modified when interactions are taken
into account. We describe the interaction by introducing in the
two-particle Schr\"{o}dinger equation the term $V({\bf x},{\bf
y})\psi ({\bf x},{\bf y})$. The potential $V$ includes both
inter-particle and particle-external field interactions. By the
matter of simplicity we assume $V$ to be real and only to depend on
the position variables of the problem.

In a straightforward way we obtain the continuity equations with
interaction:
\begin{equation}
\frac{\partial \pi _R}{\partial t} + \nabla \cdot {\bf J}_R
=-\frac{\pi _I}{\hbar} \;  {\cal U}
\end{equation}
and
\begin{equation}
\frac{\partial \pi _I}{\partial t} + \nabla \cdot {\bf J}_I
=\frac{\pi _R}{\hbar} \;  {\cal U}
\end{equation}
with
\begin{equation}
{\cal U}({\bf x},{\bf y},{\bf X},{\bf Y})=V({\bf x},{\bf y}) -V({\bf X},{\bf y}) + V({\bf X},{\bf Y})- V({\bf x},{\bf Y})
\end{equation}
The terms in the r. h. s. of the two equations represent the sources
of $\Pi $-density. They describe as the entanglement density can be
generated or destroyed. From these equations it is clear than in the
absence of interaction the $\Pi $-density can only flow.

Another important characteristic of these equations is that they
connect $\pi _R$ and $\pi _I$. During free evolution they are two
independent magnitudes. The evolution of $\pi _R$ does not depend at
all on $\pi _I$, and vice versa. In contrast, when there is
interaction the source terms of $\pi _R$ and $\pi _I$ depend
respectively on $\pi _I$ and $\pi _R$. Thus, although $\pi _I$ does
not contribute directly to $\Pi $ (Eq. (\ref{eq:doz})), it plays an
essential  role in its generation.

\section{Entanglement underlying dynamics}

The continuity equations derived in the previous sections could be
seen as a merely formal exercise. However, we shall argue that they
suggest the existence of an underlying entanglement dynamics or
entanglement subdynamics. The very possibility of formulating a
continuity equation for a physical variable indicates that it can
locally change, generating flows towards adjacent points. There is a
dynamical behavior of the variable. In our case we can say that even
for freely evolving two-particle systems the entanglement density is not a
static phenomenon but a dynamical one.

We define the entanglement subdynamics as the dynamics of the
entanglement density. We denote as the dynamical space the space
where this subdynamics takes place. The subdynamics and the dynamics
of entanglement must be compatible, suggesting that it is possible
in principle to obtain some information about the second one from
the first one. We shall discuss next two examples of such potential
applications.

A fundamental magnitude to characterize any dynamics is its
dimension, that is, the dimension of the space where it takes place.
We take here a further step and propose to measure the complexity of
the entanglement phenomenon using a parameter related to that
dimension. We introduce the Dimension Comparison Parameter (DCP) as
the ratio between the dimension of the subdynamics and the natural
dimension. The natural dimension is that of the space where the wave
functions are defined. A DCP larger that one implies that the
natural space does not provide enough mathematical structure to
describe the phenomenon and it is necessary to consider spaces with
a larger dimension. We expect the complexity of a phenomenon to
increase with the DCP.

In the case of probability the space of the probability density is
$R^6$, whose dimension coincides with that of the natural space. We
do not need to extend the natural space to describe the probability
density. In contrast, for entanglement $\pi$ is defined in $R^{12}$,
much larger than the natural space $R^6$. The DCP is two, indicating
a large complexity for the problem. We need to double the
dimensionality to accommodate the entanglement subdynamics.

This dimensionality also determines the form of the flows. In a
classical fluid the flow is between points of the physical space
($R^3$). In the case of the quantum probability the picture changes.
The two-particle probability flow no longer lies on the physical
space but in $R^6$. When we move to our problem the situation is
even more involved. The flux now is a complex variable and we must
consider two physical flows, one for each component of the current.
Moreover, each one of the physical flows lies in $R^{12}$. Each one
of the flows connects adjacent points in the $R^{12}$ space, which
correspond to four different points of the physical space (not
necessarily adjacent).

The last property suggests an interesting picture of the relation
between the global and local aspects of entanglement. The
possibility of formulating continuity equations for a system
guarantees the existence of a local dynamics in the space where they
are defined. Here, we use the term local in the sense that only
adjacent points affect the behavior of a given point at a given
time, not in the relativistic one. It can be understood as opposite
to global dynamics (we use the word global instead of non-local to
try to avoid any confusion with the use of the last term in the
context of Bell-type inequalities). For two particles this local
behavior takes place in the space $R^{12}$. However, in the physical
space there is not a continuity equation and the dynamics is global.
The absence of a continuity equation is a consequence of the fact
that the flow connects sets of four points. Our formulation provides
a picture of the relation between the global aspects of entanglement
(connecting separated points in the physical space) and the local
ones (only connecting adjacent points in the dynamical space). It
suggests an image where the entanglement is local in the dynamical
space, but with a global appearance when we operate in the physical
space. Extensions of this type of reasoning could be on the basis of
a better understanding of the global aspects of quantum theory.

\section{Discussion}

In the paper we have introduced a pair of complementary continuity
equations to describe the purity of entangled continuous variable
systems in pure states. The continuity equations, introduced for
free evolving states, remain valid in interacting scenarios after
the addition of source terms. This way we can identify the physical
form of the terms creating or destroying entanglement. We have used
the above mathematical formalism to outline a theory of the
underlying dynamics, which gives some insights on the
characterization of entanglement based on its dimension and on their
global aspects.

Our approach lies on the analogy with the description of the
probability flow via a continuity equation.  At variance with the
probability description two densities and two currents are involved
in the problem. This fact strongly suggests that entanglement is a
more complex physical phenomenon than probability. We can make this
argument more quantitative. We can use the dimension of the space of
the underlying dynamics as a measure of the complexity of the
phenomenon. As discussed before, for two particles this dimension
goes as the second power of that of the quantum probability space of
the system. In general we can take this dimension, via the DCP, as a measure of
complexity.

In \cite{Sou} it has been noted that the particular dynamics of the
entanglement present in that problem cannot be viewed as a flow of
this quantity between the parties. The same conclusion can be easily
extended to our case, where it is possible to introduce a flux or
current but neither can be interpreted it as a flow between parties.

The fact that the flow connects four different points in $R^3$, not
necessarily contiguous, is a clear manifestation of the global
character of entanglement. This global feature is not exclusive from
entanglement. For quantum probability the flow also connects points
in $R^6$, that is, pairs of points in $R^3$ not necessarily
contiguous. Global flows seem to be an inherent characteristic of
quantum multi-particle processes.

\end{document}